\documentclass[%
 aip,
 amsmath,amssymb,
 reprint,%
]{revtex4-1}

\usepackage{graphicx}
\usepackage{dcolumn}
\usepackage{bm}
\usepackage{xcolor}

\usepackage[utf8]{inputenc}
\usepackage[T1]{fontenc}
\usepackage{mathptmx}
\usepackage{etoolbox}

\makeatletter
\def\@email#1#2{%
 \endgroup
 \patchcmd{\titleblock@produce}
  {\frontmatter@RRAPformat}
  {\frontmatter@RRAPformat{\produce@RRAP{*#1\href{mailto:#2}{#2}}}\frontmatter@RRAPformat}
  {}{}
}%
\makeatother
\begin{document}

\preprint{AIP/123-QED}

\title[Predicting solvation free energies with an implicit solvent machine learning potential]{Predicting solvation free energies with an implicit solvent machine learning potential}
\author{Sebastien Röcken}
\thanks{These two authors contributed equally.}

\author{Anton F. Burnet}%
\thanks{These two authors contributed equally.}
\altaffiliation[Current address: ]{Faculty of Physics and Center for NanoScience, Department of Veterinary Sciences, Ludwig-Maximilians-Universit\"at M\"unchen, Munich, Germany}

\author{Julija Zavadlav}
\altaffiliation[Also at: ]{Atomistic Modeling Center, Munich Data Science Institute, Technical University of Munich, Germany}%
\email{julija.zavadlav@tum.de}
\affiliation{Multiscale Modeling of Fluid Materials, Department of Engineering Physics and Computation, TUM School of Engineering and Design, Technical University of Munich, Germany}

\date{\today}

\begin{abstract}
Machine learning (ML) potentials are a powerful tool in molecular modeling, enabling ab initio accuracy for comparably small computational costs. Nevertheless, all-atom simulations employing best-performing graph neural network architectures are still too expensive for applications requiring extensive sampling, such as free energy computations. Implicit solvent models could provide the necessary speed-up due to reduced degrees of freedom and faster dynamics. Here, we introduce a Solvation Free Energy Path Reweighting (ReSolv) framework to parametrize an implicit solvent ML potential for small organic molecules that accurately predicts the hydration free energy, an essential parameter in drug design and pollutant modeling. With a combination of top-down (experimental hydration free energy data) and bottom-up (ab initio data of molecules in a vacuum) learning, ReSolv bypasses the need for intractable ab initio data of molecules in explicit bulk solvent and does not have to resort to less accurate data-generating models. On the FreeSolv dataset, ReSolv achieves a mean absolute error close to average experimental uncertainty, significantly outperforming standard explicit solvent force fields. Compared to the explicit solvent ML potential, ReSolv offers a computational speedup of four orders of magnitude and attains closer agreement with experiments. The presented framework paves the way toward deep molecular models that are more accurate yet computationally cheaper than classical atomistic models.
\end{abstract}

\maketitle

\section{Introduction}\label{sec1}
Solvation free energy, and notably hydration free energy, is generally recognized as a fundamental thermodynamic quantity of interest in computational chemistry. Defined as the work done when transferring a molecule from the gas phase to the solution (water in the case of hydration free energy), it enables the computation of several key physicochemical properties of molecules, such as solubility, partition coefficients, activity coefficients, and binding free energies in solutions~\cite{fossat2021,nerenberg2012}. These properties are of great importance to the pharmaceutical, environmental, and materials sciences~\cite{ratkova2015,boobier2020,deng2020,zubatyuk2019,rauer2020,hutchinson2019solvent,Lim2019}, prompting the organization of blind prediction SAMPL challenges~\cite{nicholls2008,Geballe2010,guthrie2014} with hydration free energy as one of the main targets. In addition, Mobley et al. compiled and curated a FreeSolv database of experimentally measured hydration free energies for small neutral molecules in water~\cite{mobley2014freesolv, duarte2017approaches}.

A wide spectrum of methods is available to calculate solvation free energy, ranging from traditional approaches such as continuum solvation models~\cite{Klamt2010,Ehlert2021} to recent machine learning (ML) algorithms~\cite{Riniker2017,Zhang2023,Lim2021,Low2022,Zhang2022a,Wu2018,Yang2019,Cho2019,Pathak2021,chen2021} and their combinations~\cite{alibakhshi2021improved,Scheen2020,Weinreich2021}. The alchemical methods with Molecular Dynamics (MD) simulations~\cite{duarte2017approaches,boulanger2018optimized,karwounopoulos2023calculations} are typically assumed to be highly accurate but computationally expensive~\cite{luukkonen2020predicting,zhang2017}. However, both the fidelity and the efficiency highly depend on the explicitly treated degrees of freedom and the employed potential energy model. 

In implicit solvent models, the solvent molecules (e.g., water molecules) are not explicitly present in the system (as in explicit solvent models); instead, the interactions are modified to account for the solvent effects~\cite{onufriev2008}. The number of degrees of freedom is thereby greatly reduced, resulting in large computational gains. Classical implicit models treat the solvent as a continuum medium with specific dielectric and interfacial properties. The typical approach decouples the electrostatic (polar) and nonpolar interactions. The former can be approximated by solving the Poisson–Boltzmann equation or further simplified with the popular generalized Born model, while the latter is most often estimated via the solvent-accessible surface area~\cite{onufriev2008,roux1999}. While recent advances, such as using ML to predict the generalized Born radii~\cite{mahmoud2020}, have increased the accuracy of these models, they are still in considerable disagreement with experimental data and explicit solvent models~\cite{zhou2002,chen2021,cumberworth2016free}. In particular, the solvation free energy root mean square error (RMSE) is approximately 3.6 kcal/mol for both the Poisson-Boltzmann and Generalized Born models. By specifically optimizing the nonpolar interactions using hydration free energy data, the test set RMSE can be decreased to 1.68 kcal/mol~\cite{brieg2017generalized}.

Higher accuracy can be achieved through ab initio methods, based on quantum chemical calculations, albeit at the cost of computational demand. Commonly utilized is the COSMO family \cite{klamt1993cosmo,klamt2000cosmo}, with COSMO-RS (Conductor-like Screening Model for Real Solvents) being particularly notable in property predictions. In particular, extensions thereof have achieved high fidelity predictions of solvation free energies (MAE=0.52 kcal/mol)\cite{klamt2015calculation}. 


An alternative to the computational hurdle imposed by ab initio methods, without significantly compromising predictive accuracy, is offered by the many-body, flexible potential energy surfaces, that characterize ML potentials. During the past decade, ML potentials were deployed to derive atomistic~\cite{Friederich2021} and coarse-grained~\cite{Gkeka2020} models based on training data provided by high-fidelity simulations~\cite{ThalerRE}, experiments~\cite{thaler2021learning}, or both~\cite{Roecken2024}. Deriving empirical force fields from experimental data has been extensively studied, see e.g. Refs.~\cite{ding2024optimizing,poleto2022integration,cesari2019fitting,frohlking2020toward}, and this approach naturally extends to ML potentials. In the context of implicit water models, the capacity of deep ML potentials is large enough to compensate for the removal of solvent degrees of freedom. The predicted solute properties, such as the conformational landscape of proteins, match those obtained from the reference explicit water model~\cite{vlachas2021accelerated,Coste2023,chen2021,wang2021,wang2020,husic2020,wang2019AE,Yao2023}. However, these studies used classical atomistic force fields as the data-generating model, inherently limiting the attainable accuracy of the resulting ML potential. As shown previously, and also in this work, classical atomistic models such as General Amber Force Field (GAFF) and CHARMM General Force Field (CGenFF) systematically overestimate hydration free energies~\cite{boulanger2018optimized}. Modifying the Lennard-Jones parameters yields an improved agreement with experiments but can, at the same time, negatively affect other properties~\cite{boulanger2018optimized}.

On the other hand, more accurate ab initio training data is prohibitively expensive for solutes in explicit bulk water. Consequently, frequently used Density Functional Theory (DFT) training databases such as QM9~\cite{ramakrishnan2014}, ANI~\cite{smith2017}, and QM7-X~\cite{hoja2021qm7} contain small organic molecules in vacuum or at best contain samples with a few randomly placed water molecules around solutes~\cite{unke2019,basdogan2019}. Moreover, implicit solvent models are coarse-grained models. When trained with the common force-matching approach~\cite{Noid2008}, coarse-grained ML models require much more data than their atomistic counterparts~\cite{Durumeric2023a}. The underlying reason is the surjective atomistic-to-coarse-grained mapping, resulting in noisy force labels. As an illustration, parametrization of coarse-grained models for alanine dipeptide in implicit water required a dataset with $10^6$ configurations~\cite{Chen2021a,ThalerRE}, a number easily obtainable with classical force fields but vastly out of reach for ab initio calculations.

In this work, we present a Solvation Free Energy Path Reweighting (ReSolv) approach to parametrize an ML potential for small organic molecules in an implicit aqueous solvent. The above mentioned difficulties are circumvented with a two-stage training procedure, utilizing first the DFT database of molecules in a vacuum and then the experimental hydration free energy database. The second stage entails a non-trivial top-down training since the hydration free energy is not a direct output of an ML model but instead involves molecular simulations. Constructing the free energy integration path along the ML model training process and utilizing the Zwanzig reweighting scheme enables us to perform efficient training that avoids differentiating through the molecular simulation. The ReSolv model predicts hydration free energies more accurately than the classical explicit solvent models despite being an implicit solvent (i.e., coarse-grained) model. In addition, ReSolv's predictions are not systematically biased and are more robust for molecules with large negative hydration free energies. We also investigate error correlation between different modeling approaches and point out several potentially erroneous data points that should be reconsidered in future database curation.

\section{Methods}
\subsection{Solvation Free Energy Path Reweighting (ReSolv)}
\begin{figure*}[!htb]
\includegraphics[trim={0cm 3cm 0 0},clip,width=0.9\linewidth]{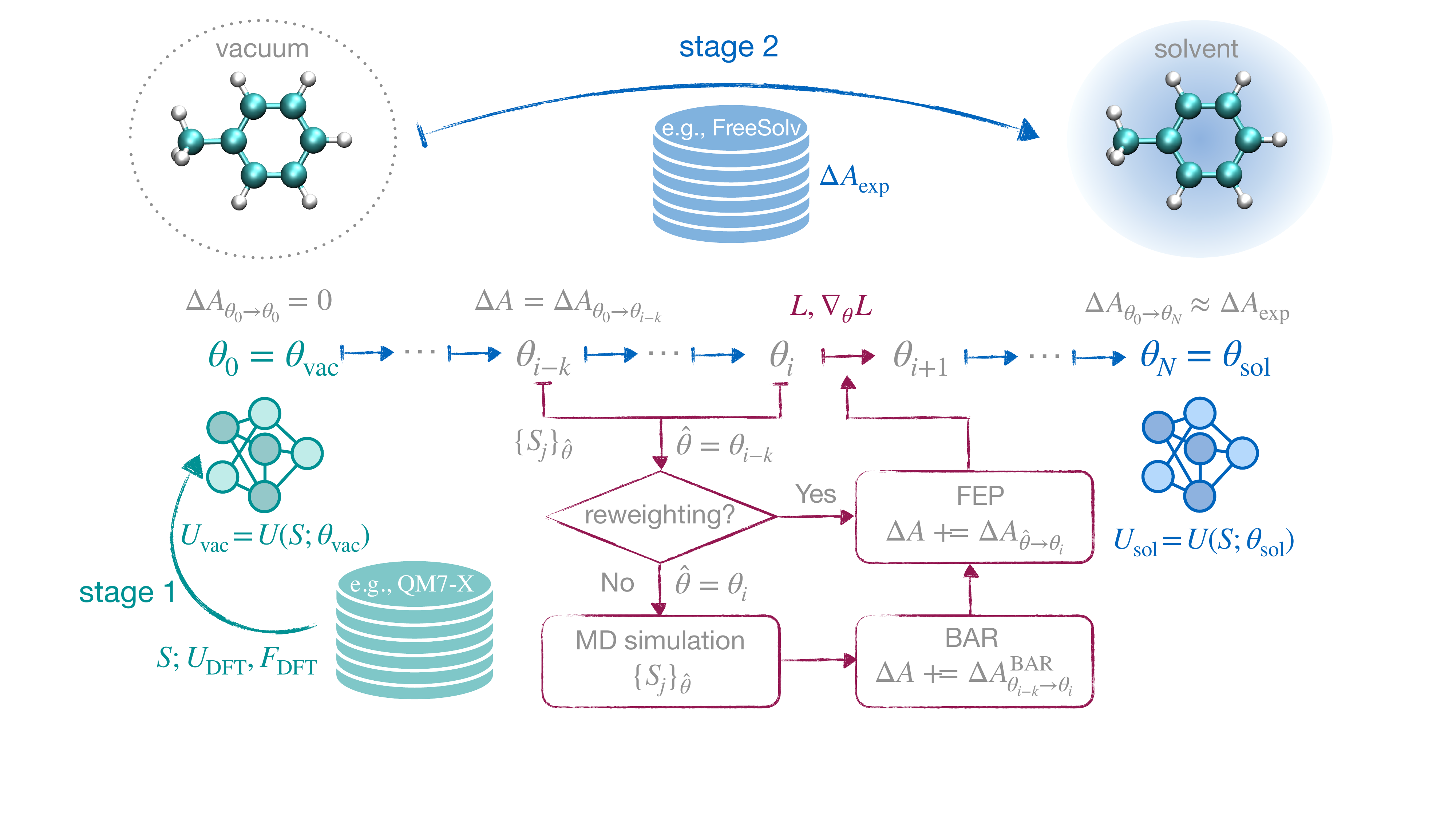}  
\caption{Solvation Free Energy Path Reweighting (ReSolv). The green color indicates ReSolv's stage one, where we train an ML potential $U_{\mathrm{vac}}$ for molecules in a vacuum based on the ab initio dataset containing configurations $S$ and the corresponding energies $U_\mathrm{DFT}$ and forces $F_\mathrm{DFT}$. The blue color represents ReSolv's stage two, where we train an ML potential $U_{\mathrm{sol}}$ for molecules in an implicit solvent by initializing the parameters with $\theta_{\mathrm{vac}}$ and perturbing them towards $\theta_{\mathrm{sol}}$ where the free energy difference between $U_\mathrm{vac}$ and $U_{\mathrm{sol}}$ equals experimental solvation free energy $\Delta A_{\mathrm{exp}}$. The red color depicts the parameter update procedure involving trajectory reweighting, Free Energy Perturbation (FEP), and Bennett acceptance ratio (BAR) methods. See main text for more details.}
\label{fig:Scheme}
\end{figure*}

The training methodology of ReSolv consists of two consecutive stages as shown in Fig.~\ref{fig:Scheme}. 
In the first stage, we parametrize the ML potential for molecules in a vacuum. The model takes as an input a configurational state of the molecule $S$ and predicts the potential energy, i.e., $U_{\mathrm{vac}}=U(S;\theta_{\mathrm{vac}})$. The forces on the atoms are computed as the negative derivative of the potential with respect to the atoms' position vectors. In training, we aim to adjust the parameters of the model $\theta_{\mathrm{vac}}$ such that the predictions match the corresponding energies $U_{\mathrm{DFT}}$ and forces $F_{\mathrm{DFT}}$ in the ab initio database, i.e., using a bottom-up training approach via Eq.~\ref{equ:lossS1}.

In the second stage, we keep the $U_{\mathrm{vac}}$ model fixed and train the ML potential $U_{\mathrm{sol}}=U(S;\theta_{\mathrm{sol}})$, parametrizing molecular interactions in an implicit solvent. The parameters $\theta_{\mathrm{sol}}$ are optimized such that the free energy difference $\Delta A$ between the potentials $U_{\mathrm{vac}}$ and $U_{\mathrm{sol}}$ reproduces the experimental solvation free energy $\Delta A_{\mathrm{exp}}$. For the sake of simplicity, we assume here that there is only one molecule in the training database. The loss function is then given by
\begin{equation}\label{eq:lossFE}
    L = \left ( \Delta A - \Delta A_{\mathrm{exp}}   \right )^2.
\end{equation}
The top-down training on experimental data is not as straightforward as bottom-up training in the first stage because $\Delta A$ is not an output of the ML model but rather evaluated from an MD simulation driven by an ML potential. We employ a variation of the Differential Trajectory Reweighting (DiffTRe) method~\cite{thaler2021learning} that avoids exploding gradients and reduces the computational and memory requirements compared to alternative gradient computation via backpropagation through the MD simulation. Before continuing, we first give a brief summary of the DiffTRe method.

Consider the task of matching a time-independent observable $O$, e.g., by postulating the following loss function $L=(\langle O \rangle_{\theta} - O_{\mathrm{exp}})^2$, where $\langle  \rangle_{\theta}$ denotes the ensemble average with respect to the canonical distribution using ML potential $U_{\theta}=U(S;\theta)$. DiffTRe leverages thermodynamic perturbation theory~\cite{zwanzig1954high}, stating that $\langle O \rangle_{\theta}$ can be estimated from $N$ decorrelated states $\{S_j\}_{\hat{\theta}}$ generated by a reference potential $U_{\hat{\theta}}=U(S;\hat{\theta})$, that is
\begin{equation}\label{eq:reweighiting}
    \langle O \rangle_{\theta} \approx \sum_{j=1}^N w_j O(S_j,U_{\theta}); \quad w_j= \frac{e^{-\beta[U(S_j;\theta)-U(S_j;\hat{\theta})]} } { \sum_{k=1}^N e^{-\beta[U(S_k;\theta)-U(S_k;\hat{\theta})]} }.
\end{equation}
where $\beta=1/(k_BT)$, $k_B$ is Boltzmann's constant and $T$ temperature. Due to limited sampling, the estimation should only be used if the states generated by the reference potential $U_{\hat{\theta}}$ are statistically close to states that would have been sampled from the potential $U_{\theta}$. The distribution overlap is captured with the effective sample size~\cite{Carmichael2012} given by
\begin{equation}\label{eq:eff}
N_{\mathrm{eff}} \approx e^{-\sum_{j} w_j \text{ln} w_j}.
\end{equation}
The DiffTRe training, therefore, works as follows. First, an MD simulation is performed with the reference potential generating $\{S_j\}_{\hat{\theta}}$. Then, at each update step, if $\theta$ is determined sufficiently close to $\hat{\theta}$, i.e., $N_{\mathrm{eff}}\geq \Bar{N}_{\mathrm{eff}}$ for a fixed effective sample size threshold $\Bar{N}_{\mathrm{eff}}$, the trajectory is reused. Otherwise, $U_{\theta}$ is set as the new reference potential, and a new reference trajectory is generated. Most importantly, for both cases, Eq.~\ref{eq:reweighiting} provides a differentiable relation between $\langle O \rangle_{\theta}$ and the model's parameters $\theta$, thereby enabling the computation of $\tfrac{\partial L}{\partial \theta}$ required for gradient-based optimization. 

For the present task of learning the solvation free energy, DiffTRe could be directly employed as presented above. Nevertheless, the free energy difference between the potentials $U_{\mathrm{vac}}$ and $U_{\mathrm{sol}}$ is typically estimated by constructing a free energy integration path, e.g., with a new potential energy function $U(\lambda) = (\lambda-1)U_{\mathrm{vac}} + \lambda U_{\mathrm{sol}}$, and estimating the free energy differences between discrete steps of $\lambda$. This approach would require as many simulations as there are $\lambda$ steps. In addition, intermediate steps would involve an MD simulation using a linear combination of two ML potentials, rendering the update step computationally expensive.

Here, we instead use the DiffTRe learning process itself as a free energy integration path. We initialize the parameters with the pre-trained $\theta_{\mathrm{vac}}$ parameters and iteratively perturb them during training towards the desired $\theta_{\mathrm{sol}}$ parameters (see Fig.~\ref{fig:Scheme}). At each update step, we compute the free energy difference between the new and previous potential using a hybrid of the free energy perturbation~\cite{zwanzig1954high} and the Bennett Acceptance Ratio (BAR) methods~\cite{bennett1976efficient}. The free energy differences are accumulated along the training, which, at the end of the training, yield the total free energy difference between $U_{\mathrm{vac}}$ and $U_{\mathrm{sol}}$ or $\Delta A$.

Explicitly, let us consider the update step from $\theta_i$ to $\theta_{i+1}$. At this point, due to reweighting, the reference trajectory may have been generated at an earlier step $i-k$ with the ML potential $U(S;\theta_{i-k})$, where $k\leq i$. Since this is the reference trajectory, $\hat{\theta}=\theta_{i-k}$. As above, we denote the states generated by the reference potential with $\{S_j\}_{\hat{\theta}}$. The corresponding free energy difference is $\Delta A=\Delta A_{\theta_0\to\theta_{i-k}}$, which was already computed at step $i-k$. Firstly, we compute $N_{\mathrm{eff}}$ (Eq.~\ref{eq:eff}). If the reweighting criterion is satisfied, the trajectory is reused. In the opposite case, we generate a new reference trajectory with the current ML potential $U(S;\theta_{i})$. Thus, $\hat{\theta}=\theta_{i}$.  We also update $\Delta A = \Delta A + \Delta A^{BAR}_{\theta_{i-k}\to\theta_i}$, where $\Delta A^{BAR}_{\theta_{i-k}\to\theta_i}$ denotes the free energy difference between steps $i-k$ and $i$ estimated with the BAR method~\cite{bennett1976efficient}. For both cases, we then compute the free energy difference between potentials $U(S;\hat{\theta})$ and $U(S;\theta_{i})$ using the differentiable free-energy perturbation relation~\cite{zwanzig1954high} 
\begin{equation}\label{equ:zwanzig}
     \Delta A_{\hat{\theta}\rightarrow \theta_i }  =  - \beta^{-1} \ln \left( N^{-1}\sum_j e^{-\beta[U(S_j;\theta_i) - U(S_j;\hat{\theta} ) ]} \right), 
\end{equation}
where the summation runs over states generated by the reference potential. Lastly, we update $\Delta A = \Delta A + \Delta A_{\hat{\theta}\rightarrow \theta_i }$ and evaluate the loss (Eq.~\ref{eq:lossFE}). Crucially, Eq.~\ref{equ:zwanzig} provides a differentiable relation between $\Delta A$ and $\theta_i$, enabling the computation of $\tfrac{\partial L}{\partial \theta_i}$. Note that if the reference trajectory was regenerated, then $\hat{\theta}=\theta_{i}$ and Eq.~\ref{equ:zwanzig} reduces to zero, but its gradient with respect to $\theta_i$ is generally non-zero. 

\subsection{Model training and validation}
Both vacuum and solvent ML potentials are based on the GNN architecture NequIP~\cite{batzner20223} as implemented in JAX MD~\cite{jaxmd2020}. Computations were performed with double precision on Nvidia A100 80GB GPUs. The architectural hyperparameters of NequIP are listed in supplementary material Table~S2. 

In the first stage, the ML potential parameters are adjusted via backpropagation so that the predicted energies and forces match the target values. The corresponding loss function is \\
\begin{widetext}
\begin{equation}\label{equ:lossS1}
    L = \frac{1}{N_{bs}} \sum_{i=1}^{N_{bs}} \left( \gamma_U (U_{i} - U_{i,\mathrm{DFT}})^2 +   \gamma_F \frac{1}{3N_{i,a}} \sum_{j=1}^{N_{i,a}} \sum_{k=1}^{3} (F_{ijk} - F_{ijk,\mathrm{DFT}})^2 \right),
\end{equation}
\end{widetext}
where $\gamma_U$ and $\gamma_F$ are weighting factors, $N_{bs}$ is the batch size, $N_{i,a}$ the number of atoms of molecule $i$, and $k$ iterates over the x-, y-, z-dimensions. $U_{i}$ is the energy of the $i$-th molecule in the batch, and $F_{ijk}$ is the force in direction k of atom j from the i-th molecule in the batch. Subscript DFT denotes the ab initio data target. We normalize the target energies and forces. The energies are shifted by the mean energy of the training data and scaled with the average root mean squared force of the training data. The forces are scaled with the average root mean square force of the training data. Further, we do not employ learnable scaling or shifting and set the per-atom scaling to 1 and the per-atom shift to 0. To pick the best model, we employ early stopping on the validation dataset. For the numerical optimization hyperparameters, see supplementary material Table~S3. 

In the second stage, we employ the RDKit~\cite{rdkit} to generate the 3D structure of a molecule based on the SMILES string provided by the FreeSolv database. Next, we perform the energy minimization of structures with the MMFM94 force field~\cite{halgren1996merck, halgren1996merck1,halgren1996merck2, halgren1996merck3, halgren1996merck4}. The obtained configurations are used to run a 300~ps initial equilibration simulation followed by a 200~ps production run for each molecule. All simulations are performed in the NVT ensemble and numerically integrated with the velocity Verlet scheme using a 1~fs timestep. To match the conditions of the experimental data the temperature is kept at 298.15~K with the Langevin thermostat with damping factor set to 1/ps. We check the stability of our simulations by evaluating that the last configuration of the molecules can be represented by a single graph and that no atoms diverged from the molecule. Additionally, to see whether our models yield physically reasonable trajectories, we compare the bonds, angles, and dihedrals of the $U_{sol}$ trajectories for three randomly selected molecules to the bonds, angles, and dihedrals as obtained after an energy minimization with the UFF force field using RDKit~\cite{rdkit}; see supplementary material Fig.~S9. In training, new simulations are initialized with the last configuration of the previous simulation and consist of 50~ps equilibration and a 200~ps production run. We sample every 5~ps during the production run, i.e., 40 samples per trajectory. The reweighting effective sample size threshold is fixed to $\Bar{N}_{\mathrm{eff}} = 0.9$. The numerical optimization hyperparameters are reported in supplementary material Table~S3. 

The reported hydration free energies are computed with the BAR method using only the end states, i.e., vacuum and water states. As previously reported, intermediate states are not necessary for implicit solvent models due to the sufficient overlap between the vacuum and solvated ensembles of the solute~\cite{zhang2017}. We first run 300~ps of equilibration and sample every 5~ps in the subsequent 200~ps production run. We perform convergence tests (supplementary material Fig.~S8) for 12 randomly selected molecules in the test set to ensure sufficient sampling.
\section{Results and discussion}\label{sec2}
\subsection{ReSolv outperforms classical explicit solvent force fields on FreeSolv database}\label{sec2a}
In this work, we employ ReSolv to learn the hydration free energy of small organic molecules, which, by construction, also yields an ML potential for molecules in implicit water. The architecture of the ML potential is based on NequIP~\cite{batzner20223}, a data efficient E(3)-equivariant graph neural network. For stage one of training, we use the QM7-X dataset~\cite{hoja2021qm7}, providing the target DFT energies and forces. The dataset consists of 4.2 million small organic molecule samples, including equilibrium states, structural isomers, structural stereoisomers, and off-equilibrium configurations. The considered molecules are composed of the heavy atom set \{C, N, O, S, Cl\}. They have up to seven heavy atoms or four to twenty-three atoms in total, including hydrogen. We randomly split the dataset into train (89.8\%), validation (10.0\%), and test (0.2\%) sets. The ReSolv vacuum model ($U_{\mathrm{vac}}$) achieves a test set mean absolute error (MAE) within literature benchmarks~\cite{duval2023faenet, unke2021spookynet} (Table~\ref{tab:MAE}). 
\begin{table}[h]
\begin{tabular}{@{}lll@{}}
\hline
 & \multicolumn{2}{c}{QM7-X test dataset MAE} \\
Model  &  Energy [eV/atom] &  Force  [eV/Å] \\
\hline
ReSolv vacuum       & \textbf{0.005}  & 0.039  \\
SchNet*              	& 0.042  & 0.056 \\
SpookyNet*           & 0.011  & \textbf{0.015} \\  
FAENET*              & 0.011  & 0.018 \\
\hline
 & \multicolumn{2}{c}{FreeSolv test dataset} \\
  & MAE [kcal/mol] & RMSE [kcal/mol]  \\
\hline
ReSolv    & \textbf{0.63}     & \textbf{0.96}   \\
Amber      & 1.02        	& 1.39  \\
CHARMM   & 1.05         & 1.79   \\  
\hline
\end{tabular}
\caption{Top: Mean absolute error (MAE) of energy and force predictions computed on 10100 random test samples in the QM7-X dataset. Results marked with an asterisk (*) were computed by Duval et al.~\cite{duval2023faenet} on a different test set. Bottom: The MAE and root mean square error (RMSE) of hydration free energy prediction of different models for the same test set molecules in the FreeSolv database. The best-performing models are highlighted in bold.}
\label{tab:MAE}
\end{table}
The errors are close to the QM7-X dataset error, i.e., the DFT calculations were computed at the PBE0+MBD level, with a precision of $10^{-3}$ eV and $10^{-4}$ eV/Å for the energies and forces, respectively. 

For stage two of training, we utilize the FreeSolv database~\cite{duarte2017approaches}. It contains hydration free energies of 643 small molecules with neutral charges, representing compounds relevant to drug-like molecules. Keeping consistent with the heavy atom types of the QM7-X dataset, i.e., \{C, N, O, S, Cl\}, we extract the corresponding subset with 559 molecules. We also excluded seventeen molecules for which the ReSolv vacuum model yielded unstable simulations and an additional five molecules that were unstable during training. We split this subgroup into train and test sets of roughly 70\% and 30\% proportions, which amounts to 375 and 162 molecules, respectively. When dividing the data, we kept the occurrences of various heavy atom combinations consistent within the two sets, i.e., the training set contains roughly 70\% of the molecules with heavy atom combinations N, N-C, N-C-O, etc. This choice ensures that the various heavy atom combinations are represented during training and testing. In addition, we ensured that the test set consists exclusively of molecules with chemical functional groups that appear in the training set, i.e., any molecule that is a lone representative of a functional group formed part of the training set. The functional group classification was taken from the FreeSolv database~\cite{mobley2014freesolv}. The polyfunctional molecules were classified according to the first listed functional group as in Ref.~\cite{karwounopoulos2023calculations}. Apart from satisfying these two conditions, the molecules are split randomly. 

The ReSolv's test set errors are reported in Table~\ref{tab:MAE} and compared with the explicit solvent classical atomistic models Amber (GAFF) and CHARMM (CGenFF). We compute the errors for the same test set molecules to enable a direct comparison. Data for Amber is provided in the FreeSolv database~\cite{mobley2014freesolv}, while for CHARMM, the values are taken from Ref.~\cite{karwounopoulos2023calculations}. In the CHARMM error computation, two molecules (Mobley ID 6359135 and 2146331) are excluded because data is not provided in Ref.~\cite{karwounopoulos2023calculations} since the force field parameters could not be generated. The ReSolv's MAE is close to the average uncertainty of the FreeSolv database (0.6~kcal/mol). Moreover, ReSolv outperforms the explicit solvent classical atomistic models by a large margin. Amber and CHARMM models display similar MAE/RMSE for the molecules in the training set (supplementary material Fig.~S1), confirming that the test set is representative of the entire considered database. 
In addition, we trained the ReSolv model using a random splitting of the data and achieved similar performance. In particular, the obtained MAE and RMSE are 0.65 and 0.91 kcal/mol, respectively. The corresponding errors for the classical force fields remain substantially higher (supplementary material Fig.~S2). The comparison with the Amber and CHARMM models is somewhat unfair, given that the two models were not explicitly parametrized to reproduce the hydration free energy data. Nevertheless, reparametrizations and corrections in this direction were previously attempted, and the resulting models still scored lower than ReSolv. For example, Boulanger et al. rescaled the Amber (GAFF) molecule-water van der Waals dispersion interaction to better reproduce the hydration free energy and reported an MAE of 0.79~kcal/mol for the final optimized model~\cite{boulanger2018optimized}. In another example, Scheen et al. trained an ML model to correct the Amber (GAFF) force field predictions~\cite{Scheen2020}. The hybrid Amber/ML approach achieved an MAE of 0.76 kcal/mol on the SAMPL4 test data containing 47 samples of the FreeSolv database.   

\begin{figure*}[!htb]
\centering
\includegraphics[trim={0cm 0cm 0 0},clip,width=0.9\linewidth]{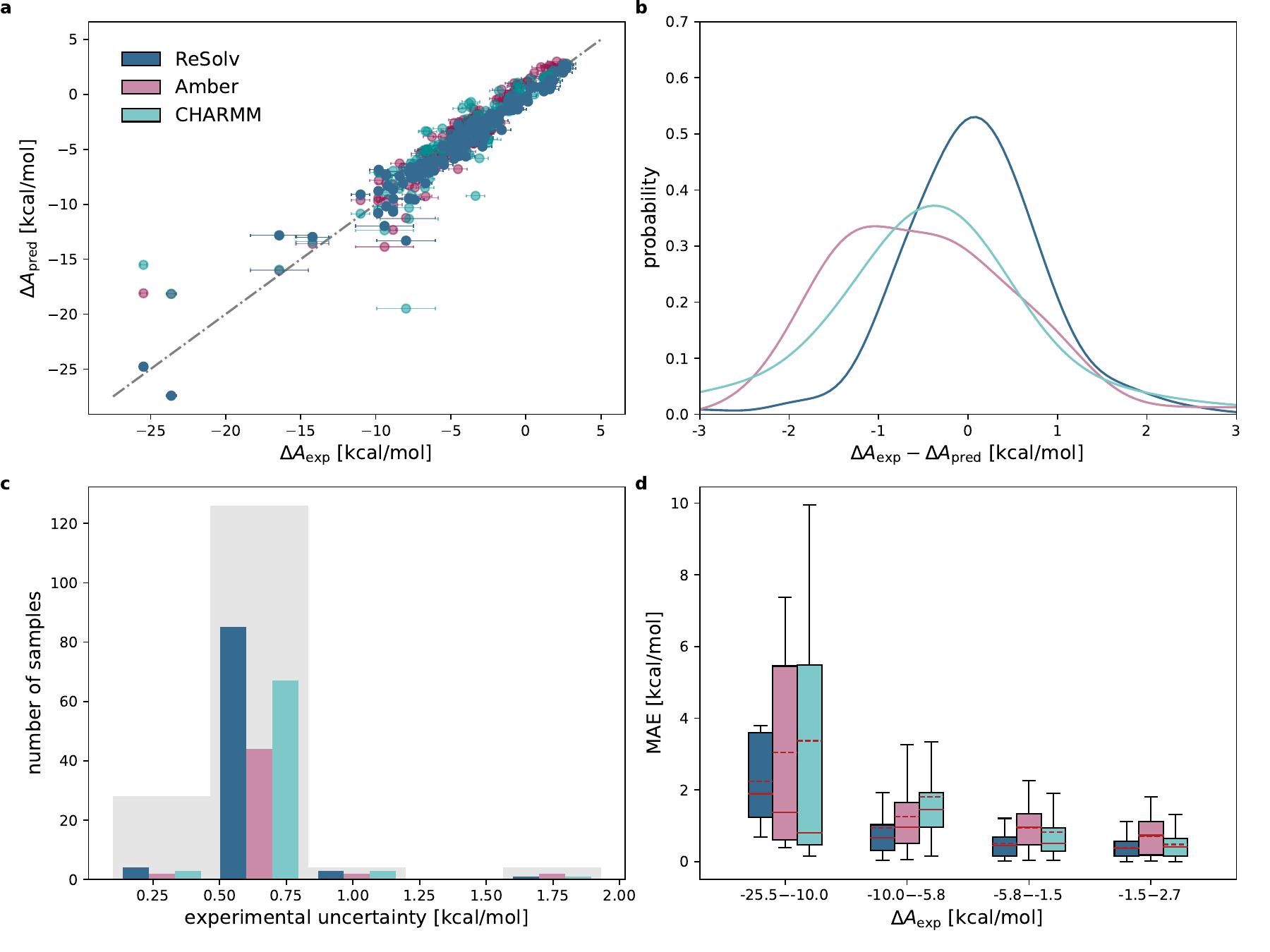}  
\caption{Prediction performance. The implicit solvent ReSolv model (blue) is referenced against the explicit solvent classical atomistic models Amber (GAFF; red) and CHARMM (CGenFF; green). The results are shown for the test dataset. (a) Parity plot with the error bars denoting the experimental uncertainty and the gray dash-dotted line indicating the perfect prediction. (b) Error probability distribution relative to the experimental values. The distributions are fitted with the Gaussian kernel density estimator. (c) The number of predictions with errors lower than the experimental uncertainty. The gray background histogram depicts the distribution of experimental uncertainty. The total percentage of predictions within the experimental uncertainty is 57\%, 31\%, and 46\% for the ReSolv, Amber, and CHARMM models, respectively. (d) Mean absolute error (MAE) increase with decreasing experimental hydration free energy. The red solid line denotes the median, red dashed line mean, the box ranges are from the first to the third quartile, and the whiskers correspond to the 1.5x interquartile range.}
\label{fig:gnnvsgff}
\end{figure*}

Next, we perform an in-depth error analysis. The parity plot (Fig.~\ref{fig:gnnvsgff}a) reveals systematic errors for Amber and CHARMM predictions, with most points lying above the diagonal line. 
To further demonstrate this point we plot the error distributions relative to the experimental values (Fig.~\ref{fig:gnnvsgff}b). The distributions are skewed to the left for classical force fields, indicating that the majority of predictions are overestimated. The Amber model, for example, overestimates the hydration free energies on average by more than 1~kcal/mol as already reported by Boulanger et al.~\cite{boulanger2018optimized}. Conversely, no systematic errors are found for the ReSolv model, evidenced by the symmetric distribution about zero. These findings are reflected in the percentage of predictions within the experimental uncertainty (Fig.~\ref{fig:gnnvsgff}c). The ReSolv model scores the highest, followed by CHARMM and Amber in last place, albeit Amber exhibiting lower MAE and RMSE than the CHARMM model. 

The parity plot (Fig.~\ref{fig:gnnvsgff}a) also highlights the FreeSolv's uneven distribution with respect to the hydration free energy value. There are only a few molecules with large and negative hydration free energy. The largest two (Mobley ID 9534740 and D-mannitol) are particularly problematic for the Amber and CHARMM models. These molecules, discussed further in the next section, are part of a general trend. The MAE increases with decreasing hydration free energy (Fig.~\ref{fig:gnnvsgff}d) for all models. However, ReSolv model is the most robust model in this respect. See supplementary material Fig.~S3 for point-wise results and supplementary material Fig.~S1 for the training dataset results. For Amber and CHARMM models, a possible explanation could be an inaccurate partial charge assignment. As shown by Jämbeck et al.~\cite{jambeck2013partial}, the hydration free energy is very sensitive to the choice of charge computation method and can differ by several kcal/mol depending on the method used. We found that the absolute sum of the Gasteiger partial charge per atom is highly correlated with the hydration free energy as is the polar surface area of the molecules (supplementary material Fig.~S4). Indeed, keeping in mind the net-neutrality of the molecules, the absolute Gasteiger partial charge per atom is sensitive to local charge gradients contributed by polar fragments within the molecule, whence its correlation with polar surface area. The computed Pearson correlation coefficient is r=-0.71, which is comparable to the correlation with the related polar surface area (r=-0.74) but significantly larger than correlation with other molecular properties such as dipole moment (r=-0.42) or volume of a solute (r=-0.18). In particular, we find that the absolute sum of the charges increases with decreasing hydration free energy. An accurate partial charge assignment is, therefore, particularly important for molecules with large negative hydration free energies, potentially explaining why the predictions for classical force fields tend to worsen with decreasing hydration free energy.
With this insight, we depict in supplementary material Fig.~S5 the difference between the $U_{\mathrm{sol}}$ and $U_{\mathrm{vac}}$ atomic contributions to the potential energy for a fixed configuration and find a correspondence with the Gasteiger partial charges from the heavy atoms. The energy differences highlight that the changes are predominant around polar atoms, agreeing with what we would expect based on the above correlations, supporting that the learning of $U_{\mathrm{sol}}$ is physically informed.

\begin{figure*}[!htb]
\includegraphics[width=0.9\textwidth]{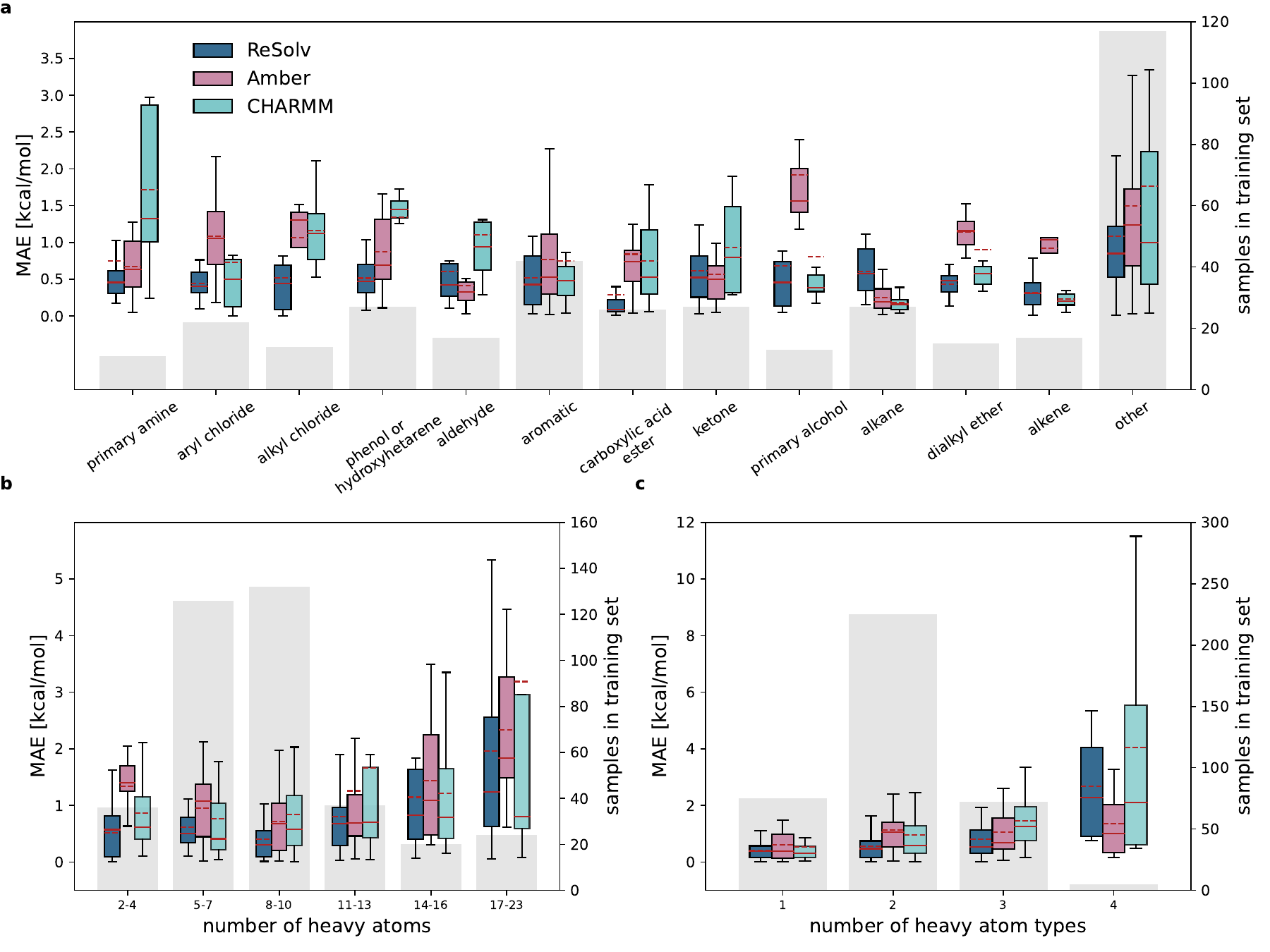}
    \caption{Error analysis. The box plots of the mean absolute error (MAE) with respect to (a) chemical functional groups, (b) number of heavy atoms, and (c) number of heavy atom types are shown for the test set molecules. The red line denotes the median, red dashed line mean, the box ranges are from the first to the third quartile, and the whiskers correspond to the 1.5x interquartile range. The gray bar plots depict the number of corresponding samples in the training set. We compare three different models: ReSolv (blue), Amber (GAFF; red), and CHARMM (CGenFF; green). In subplot (a), the `other' group corresponds to the remaining functional groups with less than three samples in the test set. }
    \label{fig:analysis}
\end{figure*}

In Fig.~\ref{fig:analysis}, the test set MAE is broken down into various molecular properties. 
First, we examine the error with respect to chemical functional groups. The chemical space covered by the FreeSolv database is quite extensive given the relatively small size of the database~\cite{karwounopoulos2023calculations,mobley2014freesolv,duarte2017approaches}. ReSolv demonstrates uniform error across the chemical functionalities, which is, to some extent, expected given the training set construction. However, low MAE is also found for the functional groups rarely appearing in the training set, i.e., for the `other' group where we merge functional groups with sparse occurrence in the train and test sets. In contrast, Amber and CHARMM models exhibit larger fluctuations in MAE for some functional groups (e.g., primary alcohol for Amber or primary amine for CHARMM) displaying notably higher errors. Note that the same conclusion can be drawn for the molecules in the training set (supplementary material Fig.~S6). Next, we consider the models' robustness with respect to the size of the molecule. From 14 heavy atoms onwards, an increased MAE can be seen for all models, most notably for the Amber model. Lastly, in terms of heavy atom types, all three predictive models display larger errors for molecules with four heavy atom types. However, we attribute this result mainly to the small sample size containing problematic molecules, i.e., the MAE is computed on only four molecules with the nitralin molecule (discussed in the next section) contributing most to the error. The absence of a similar trend for the molecules in the training set (supplementary material Fig.~S6) further supports our claim.

\subsection{ReSolv generalizes well to unseen functional groups}
The results thus far demonstrate ReSolv's ability to accurately predict the hydration free energies of molecules with chemical functional groups seen during training. Given the typically limited availability of training samples, it is especially desirable for ML potentials to extrapolate effectively into unseen chemical space. To assess ReSolv's chemical generalizability, we employ the same training procedure and dataset prescribed in Sect.~\ref{sec2a}, but with all samples of a given functional group (including multifunctional molecules) removed from the training set and compare the performance to the cases when seen.

\begin{figure*}[!htb]
\centering
\includegraphics[width=.9\textwidth]{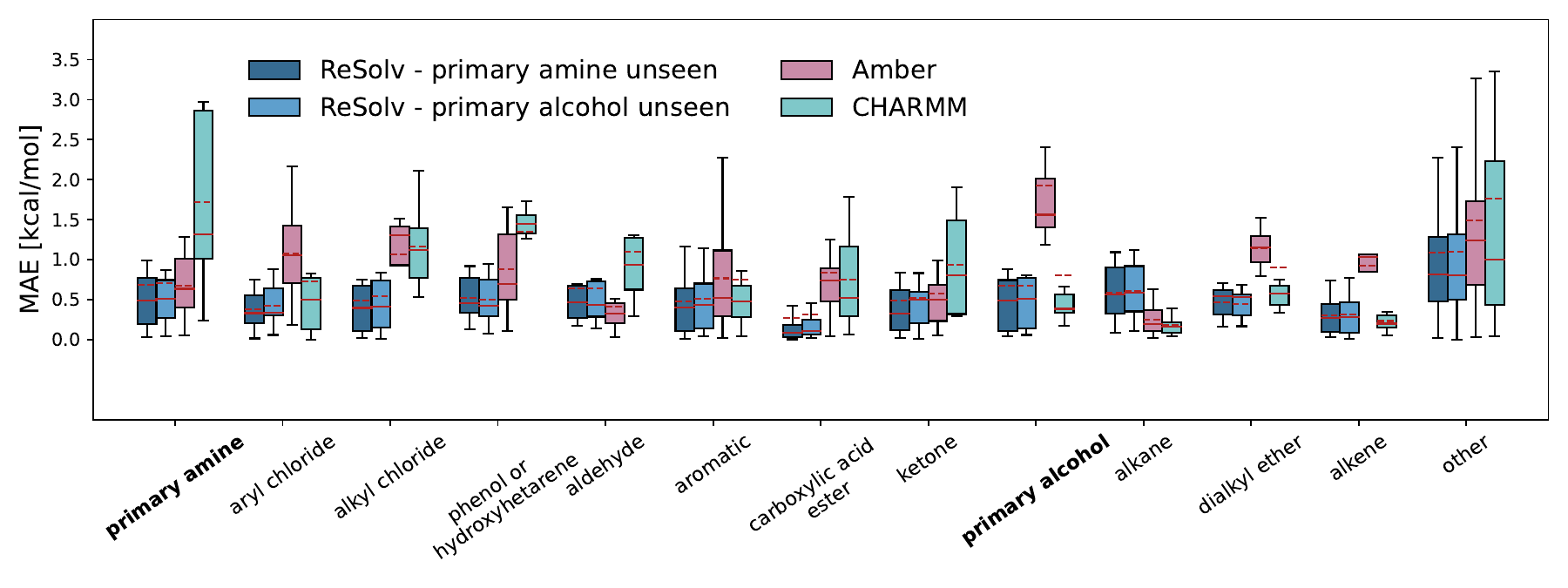}
    \caption{Extrapolation to unseen functional groups. Box plots of the mean absolute error (MAE) with respect to chemical functional groups, comparing four different models: ReSolv with primary amines unseen during training (dark blue); ReSolv with primary alcohols unseen during training (light blue); Amber (GAFF; red); and CHARMM (CGenFF; green). The `other' group corresponds to the remaining functional groups with less than three samples in the test set. The red line denotes the median, red dashed line mean, the box ranges are from the first to the third quartile, and the whiskers correspond to the 1.5x interquartile range. }
    \label{fig:unseen}
\end{figure*}

We conducted two such trainings, excluding samples with either primary amine or primary alcohol functionalities. These functional groups were selected due to their challenging nature for CHARMM and Amber, respectively (see Fig.~\ref{fig:analysis}a). With the test sets identical, we found that the errors associated with the unseen functional groups were highly consistent with those observed when the same groups were seen in training, see Fig.~\ref{fig:unseen}. This consistency demonstrates ReSolv's robustness and capacity to predict well even for unencountered regions in chemical space.

\subsection{Error correlations}
We noticed several outliers in the parity plot (Fig.~\ref{fig:gnnvsgff}a) for which the predictions are largely off for all three investigated models. Error correlation between different modeling approaches can be used to identify possible erroneous data points. Previous outlier analysis stimulated corrections of some experimental data points that are part of the FreeSolv database~\cite{Reinisch2014}. For example, the initially provided experimental value for D-mannitol was -27.79~kcal/mol which was later corrected to -23.62~kcal/mol. Nevertheless, doubts about the validity of the experimental value remained as the COSMO-RS prediction deviated by 6 kcal/mol~\cite{Reinisch2014}. As a result, some recent studies excluded D-mannitol from the test set~\cite{Zhang2023,Riniker2017}. Interestingly, ReSolv's prediction of D-mannitol's hydration free energy is -27.42 kcal/mol, in excellent agreement with the original experimental value. 

\begin{figure*}[!htb]
\centering
\includegraphics[width=.8\textwidth]{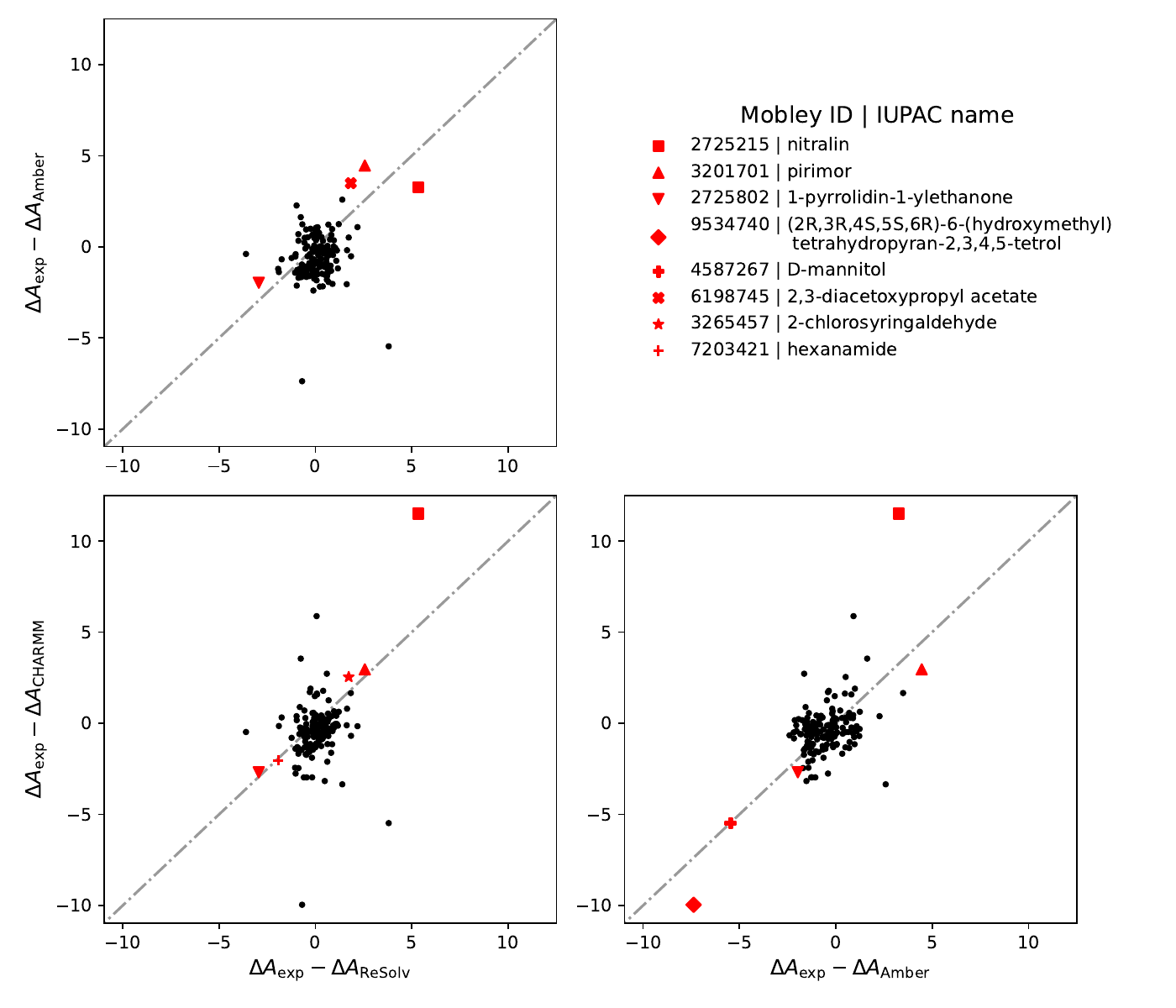}
    \caption{Correlation of errors in kcal/mol between the ReSolv, Amber (GAFF), and CHARMM (CGenFF) models for the test dataset. The gray dash-dotted line indicates the perfect correlation. Eight molecules with absolute errors greater than 1.7 kcal/mol simultaneously for x- and y-axes are marked with distinct symbols as shown in the legend. In supplementary material Table~1 we list the mean errors across all models and experimental uncertainties for these eight molecules.}
    \label{fig:correlation}
\end{figure*}

We compute the Pearson correlation coefficient of signed and absolute errors to investigate the error correlation between ReSolv, Amber, and CHARMM predictions (supplementary material Fig.~S7). Overall, we find positive correlations between models, with the largest correlation of 0.6 between the Amber and CHARMM models. In Fig.~\ref{fig:correlation}, we show point-wise error correlations and mark the molecules with large and correlated errors (exact values are given in supplementary material Table~S1). 
The nitralin and pirimor molecules particularly stand out, with all models' predictions deviating from experimental values by more than 2.5~kcal/mol. These two molecules also have a large experimental uncertainty (1.93~kcal/mol), confirming that error correlation analysis is a useful approach for inaccurate data recognition.

\subsection{ReSolv achieves four orders of magnitude speedup compared to the explicit solvent ML potentials}
Lastly, we compare the ReSolv's performance to explicit solvent ML potentials. Using the same ML potential architecture, implementation, and common simulation setup in the literature, the measured computational cost differs by four orders of magnitude (supplementary material). The ReSolv's speed up is due to due to two factors: (i) faster execution of MD step  due to reduced number of particles, (ii) reduced number of required MD steps due to enhanced sampling.

\begin{figure}[h!]
\centering
\includegraphics[trim={0cm 0cm 0 0},clip,width=0.49\textwidth]{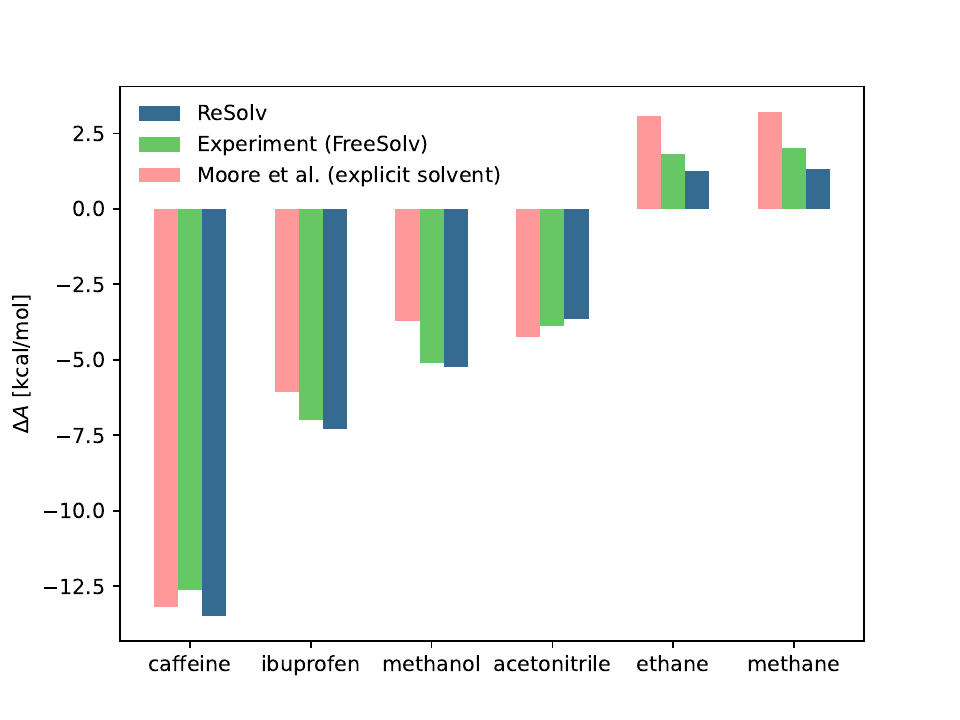}  
\caption{Comparison with an explicit solvent ML potential. The hydration free energy prediction for six molecules in the FreeSolv database recently tested by Moore et al.~\cite{moore2024computing} with an explicit solvent ML potential. Apart from caffeine, ReSolv's predictions are closer to experimental results. This includes ethane which is part of our test dataset.}
\label{fig:comapre_HFE_prediction}
\end{figure}

Surprisingly, ReSolv's substantial computational gains do not come at the cost of accuracy. To the contrary, comparison with the recent work~\cite{moore2024computing} on hydration free energy predictions for six selected molecules in the FreeSolv database suggests that the ReSolv's predictions are in better agreement with the experimental results (Fig.~\ref{fig:comapre_HFE_prediction}). More precisely, ReSolv model achieves better accuracy for five out of six molecules, including ethane, which is part of our test dataset, while the other five molecules were included in the training dataset.

\section{Conclusion}
This work presents the Solvation Free Energy Path Reweighting (ReSolv) to efficiently learn an ML-based potential energy surface in an implicit solvent by utilizing experimental solvation free energy data and reweighting in training. Since the framework was showcased with QM7-X and FreeSolv datasets, the trained ReSolv model enables an accurate hydration free energy prediction for small organic compounds, which can be directly utilized in, e.g., de novo drug design. However, the spanned chemical space of molecules could be enlarged with other datasets. In addition, the same methodology can be used to obtain an ML potential for implicit solvents other than water. For instance, extensively available octanol-water partition data could be used to derive a model in an implicit octanol environment. Predicting lipophilicity, correlated with oil-water partition coefficient, is also a critical parameter in the pharmaceutical industry as drug candidates must be sufficiently lipophilic to penetrate the lipid core of membranes but not too lipophilic that they remain there~\cite{van2004}. Transfer learning strategy would likely accelerate the learning of other solvents. It is reasonable to assume that similar solvents modify the interactions similarly. For example, the knowledge gained by learning an implicit water model could benefit the model for other polar solvents, such as ethanol. This setting could be particularly useful for solvents for which solvation free energy data is scarce.

For the FreeSolv database, ReSolv's MAE is close to experimental uncertainty. The obtained accuracy is also similar to the previously reported structure-property models, employing ML algorithms to directly predict hydration free energy from the physics-inspired fingerprints~\cite{Riniker2017,hutchinson2019solvent,Weinreich2021,Zhang2023} or molecular structure~\cite{Lim2021,Low2022,chen2021,Pathak2021, Wu2018,Yang2019,Lim2019,Cho2019}. The latter employ graph-based neural network architectures and typically perform better than the former with the reported MAE in the range of 0.58-0.76 kcal/mol and RMSE in the range of 0.82-1.23 kcal/mol. This results indicate that further improvements will likely require an enlarged and improved experimental database rather than a better modeling approach or neural network architecture. In line with this conclusion is a recent study~\cite{Zhang2022a} achieving a notably lower MAE of 0.42 kcal/mol where a graph neural network was first pre-trained on a large dataset and later fine-tuned on the FreeSolv database, i.e., by exploiting the transfer learning approach. ReSolv showcases good generalization across the functional space but a decreased generalization to larger molecules, which should be considered when planning future experiments. In addition, we found error correlations between our models and classical atomistic models that were not trained on the FreeSolv database. Molecules with large and correlated errors, such as nitralin and pirimor, call for a reevaluation of the experimental data. These points should be excluded when training future models to avoid detrimental effects.

In a broader context, ReSolv could be a first step towards a general implicit solvent ML potential with better accuracy and efficiency than classical atomistic models. Solvation free energy encodes solute-solvent interactions, which in turn govern many biomolecular processes, including folding, aggregation, and ligand binding. In vivo, these processes occur in an aqueous solution rendering the hydration free energy a general benchmark for classical force field validation~\cite{duarte2019, nicholls2008, fossat2021}, fine-tuning~\cite{huang2017, nerenberg2012}, comparison~\cite{kashefolgheta2020,shirts2003}, and calibration~\cite{oostenbrink2004, kashefolgheta2017}. Nevertheless, before using the ReSolv model in general simulations, the model needs to be tested and potentially additionally trained on other available experimental data, which we leave for future studies. Similar as for other implicit solvent ML potentials, using a prior potential (i.e., a fixed simple potential) is likely also necessary to achieve stable simulations of large macromolecules~\cite{Durumeric2023a}. Concerning computational efficiency, the ReSolv and ML-based implicit solvent models, in general, provide a speed-up compared to the classical explicit solvent force fields due to reduced degrees of freedom and associated accelerated dynamics~\cite{meinel2020loss, katzberger2024general}, even though evaluating an ML potential requires much more operations than an empirical potential~\cite{charron2023navigating}. Further gains could be obtained with constrained dynamics or a coarse-grained representation of solute molecules, enabling larger integration timesteps.  

\section*{Supplementary material}
See the supplementary material for hyperparameters and additional results including performance and error analysis on train dataset, absolute error vs experimental hydration free energy, correlation of molecular properties with experimental hydration free energy, potential energy difference vs Gasteiger charges, Pearson correlation of errors, convergence of the hydration free energy computation. 

\section*{Author contributions}
J.Z. conceptualized the study. S.R. and A.F.B. contributed equally to this work. S.R. and A.F.B. implemented and applied the ReSolv method and conducted simulations and postprocessing. All authors planned the study, analyzed and interpreted the results, and wrote the paper.

\section*{Conflicts of interest}
There are no conflicts to declare.

\begin{acknowledgments}
Funded by the European Union. Views and opinions expressed are, however, those of the author(s) only and do not necessarily reflect those of the European Union or the European Research Council Executive Agency. Neither the European Union nor the granting authority can be held responsible for them. Funded by the European Research Council (ERC) StG under Grant No. 101077842—SupraModel. The authors thank Paul Fuchs for implementing the BAR method into the chemtrain library https://github.com/tummfm/chemtrain and for insightful discussions on theory and implementation of the ReSolv method.
\end{acknowledgments}

\section*{DATA AVAILABILITY STATEMENT}
The data that support the findings of this study are available within the
article and its supplementary material and will be made openly available at \href{https://github.com/tummfm/ReSolv}{https://github.com/tummfm/ReSolv} upon acceptance of the paper. The QM7-X dataset by Hoja et al. \cite{hoja2021qm7} is available at \href{https://doi.org/10.5281/zenodo.4288677}{https://doi.org/10.5281/zenodo.4288677}. The FreeSolv dataset by Mobley and Guthrie \cite{mobley2014freesolv} is available at \href{https://escholarship.org/uc/item/6sd403pz}{https://escholarship.org/uc/item/6sd403pz}. 
The code will be made publicly available at \href{https://github.com/tummfm/ReSolv.git}{https://github.com/tummfm/ReSolv.git} upon acceptance of the paper.

\bibliography{bibliography}

\end{document}